# Quantum well stabilized point defect spin qubits

V. Ivády[1, 2], J. Davidsson[2], N. Delegan[3,4], A. L. Falk[5, 6], P. V. Klimov[5], S. J. Whiteley[5], S. O. Hruszkewycz[4], M. V. Holt[7], F. J. Heremans[3,4,5], N. T. Son[2], D. D. Awschalom[3,4,5], I. A. Abrikosov[2, 8], and Adam Gali[1, 9, *]

[1]Wigner Research Centre for Physics, Hungarian Academy of Sciences, PO Box 49, H-1525, Budapest, Hungary.

[2]Department of Physics, Chemistry and Biology, Linköping University, SE-581 83 Linköping, Sweden.

[3]Center for Molecular Engineering, Argonne National Laboratory, Lemont, IL USA.

[4]Materials Science Division, Argonne National Laboratory, Lemont, IL USA.

[5]Pritzker School of Molecular Engineering, University of Chicago, Chicago, IL, USA.

[6]IBM T.J. Watson Research Center, Yorktown Heights, NY, USA.

[7]Center for Nanoscale Materials, Argonne National Laboratory, Lemont, IL, USA

[8]Materials Modeling and Development Laboratory, National University of Science and Technology 'MISIS', 119049 Moscow, Russia.

[9]Department of Atomic Physics, Budapest University of Technology and Economics, Budafoki út 8., H-1111 Budapest, Hungary.

*Corresponding author, email: gali.adam@wigner.mta.hu

**Abstract**

Defect-based quantum systems in in wide bandgap semiconductors are strong candidates for scalable quantum-information technologies. However, these systems are often complicated by charge-state instabilities and interference by phonons, which can diminish spin-initialization fidelities and limit room-temperature operation. Here, we identify a pathway around these drawbacks by showing that an engineered quantum well can stabilize the charge state of a qubit. Using density-functional theory and experimental synchrotron x-ray diffraction studies, we construct a model for previously unattributed point defect centers in silicon carbide (SiC) as a near-stacking fault axial divacancy and show how this model explains these defect's robustness against photoionization and room temperature stability. These results provide a materials-based solution to the optical instability of color centers in semiconductors, paving the way for the development of robust single-photon sources and spin qubits.

**Introduction**

Solid-state defects have emerged as candidate systems for quantum-information technologies related to nanoscale sensing and communication [1–11], building on the ability to engineer their crystallographic environments. The sensitivity and robustness of these systems are strongly dictated by their local environment influencing the charge state stability throughout their optical initialization and read-out cycles and spin coherence times [12,13]. However, optical manipulation regularly results in undesired ionization processes, fluctuating the defect's charge state and functionality (see Fig. 1A-C) [14–19]. These undesired non-radiative pathways are a recurring challenge with most point defect qubits, including the nitrogen-vacancy (NV) center in diamond [14,15,20,21], the divacancy (VV) in silicon carbide (SiC) [7,8,16–18,22], and the silicon vacancy ($V_{Si}$) in SiC [9,16,23].

Here, we outline an engineerable method to circumvent such undesired ionization effects based on the manipulation of the host material's local band structure. Specifically, we explore how a local change in the crystal lattice stacking order (stacking fault) creates a quantum well that lowers the ionization energy of a point defect's dark state. This energy reduction, in turn, enables new optoelectronic transitions, allowing incident illumination to preferentially repopulate the desired point defect's bright state.

We elaborate on this in the context of a SiC system where a quantum well, defined by a Frank-type stacking fault[24], is predicted to stabilize divacancies in their neutral charge-state under illumination. We also provide complementary experimental evidence that such crystallographic configurations exist in commercially available SiC crystal substrates and associate the previously observed unattributed (PL5-PL7) defects in SiC[7,25] to near-stacking fault divacancies predicted to form within the well's vicinity. Unlike their bulk divacancy counterparts, these near-stacking fault divacancy defects demonstrate robustness against photoionization effects[16] and optical spin contrast persisting up to room temperature [7,25,26]. The approach highlighted in this paper provides a materials-based solution to the optical instability problem of color centers in a variety of solid state systems. This is achieved through local variation of the crystallographic environment of point defects, paving the way towards robust, defect based single photon emitters and spin qubits.

**Results**

It has been suggested that quantum well structures in wide bandgap semiconductors allow for color center defects with properties distinct from their homologues in unperturbed bulk-like crystal environments.[27] If true, these local transition energy disturbances should allow for the reversal of undesired laser induced bright-to-dark charge state effects by opening new dark-state ionization pathways. To study this, we model 4H-SiC (see Fig. 2A) as a prototypical host material with known crystallographically defined quantum wells[28–31] and color centers[7,9,32]. In this context, we investigate stacking faults (*i.e.* polytypic inclusions in 4H-SiC) as an intrinsic quantum well forming mechanism[33–35], and the effects that these structures have on divacancy defect (two adjacent empty atomic sites) color centers in their vicinity.

The insertion of a single cubic SiC double layer (see Fig. 2C) into the 4H-SiC primitive cell results in a Frank-type stacking fault (a 1FSF(3,2) fault in the Zhdanov notation) [24] as shown in Fig. 2D. The resulting stacking irregularity is seen to be indistinguishable from a polytypic 6H-SiC (see Fig. 2B) inclusion in a 4H-SiC crystal[35,36]. It has been shown that such stacking faults form quantum-well-like states that has been observed through photoluminescence (PL)[33–35] measurements in which the 6H in 4H polytypic inclusion was typically identified using the 482 nm PL-emission line[35].

Building on these observations, we calculate the change in 4H-SiC band structure with a 6H polytypic inclusion (see Fig. 1E). This stacking fault gives rise to energy states below the 4H-SiC conduction band minimum, effectively lowering the band-gap value ($E_g$) by ~0.24eV in its vicinity (up to ~15 Å from the stacking fault). This calculation does not account for electron-hole interactions that may further reduce the local $E_g$ of the defective 4H-SiC host. Nevertheless, these results illustrate that even a single stacking fault can significantly change the local $E_g$ structure. This local change in accessible energy levels, and the changes in photoionization energies that occur as a result, influence the charge state stability of local point defects states[16].

The effect of this quantum well structure on the different possible 4H-SiC divacancy sites was investigated. There are two inequivalent sites for both Si and C in the primitive cell of 4H-SiC (see Fig. 2A). Consequently, point defects have multiple nonequivalent configurations that differ primarily in the makeup of their second and further neighbor shells. The configurations of a single site defect in the 4H polytype are marked by *h* or *k*, referring to a more hexagonal or cubic-like environment, respectively (see Fig. 2A-D).

This distinction leads to divacancy defects (VV) having altogether four possible configurations in 4H-SiC: *hh*, *kk*, *hk*, and *kh*. Additionally, depending on whether the $V_{Si}$-$V_C$ axis is parallel or inclined (at ~109.5° angle) to the *c*-axis, the divacancies can exhibit either high ($C_{3v}$) or low ($C_{1h}$) symmetry. Hereinafter, we refer to these configurations as axial and basal plane divacancies respectively. In recent years, each of these photoluminescent VV configurations have been assigned to various qubits in 4H-SiC[7,37,38].

In our calculations, we considered two sets of divacancy configurations within a single model: (a) divacancies in the near-stacking fault region, up to 5 Å away from the stacking fault; and (b) divacancy configurations in a more bulk-like environment, at least 14 Å away from the stacking fault. The near-stacking fault and bulk-like configurations are denoted with a suffix –ssf (single stacking fault) and -4H, respectively (see Fig. 2D) with the prefix representing silicon and carbon vacancies in both hexagonal-like and cubic-like environment. In the context of the stacking fault, we distinguish three cubic-like and two hexagonal-like lattice sites: $k_1$, $k_2$, $k_3$, $h_1$, and $h_2$ (see colored area of Fig. 2D).

To characterize these local configurations of interest, we calculated the hyperfine and zero-field splitting (ZFS) constants for the non-equivalent divacancy sites using density functional theory (DFT). When possible, these values are paired with experimental measurements of these quantities from ref. [39]. Both the theoretical and experimental results are summarized for axial and basal divacancy configurations in Fig. 3B-F (also in Table S1-4). From this, five divacancy groups could be identified based on their splitting constants, three (1-3) for the axial and two (4-5) for the basal plane configurations.

1. The hexagonal axial configurations (*hh*-4H, $h_1h_1$-ssf, $h_2h_2$-ssf)
2. The majority of cubic-like configurations (*kk*-4H, $k_1k_1$-ssf, and $k_3k_3$-ssf) showing similar splitting within their respective groups to group 1.
3. The cubic-like $k_2k_2$-ssf divacancy presents an outlier splitting value compared to groups 1 and 2. This configuration forms in the stacking fault itself as shown in Fig. 2D.

Meanwhile, the basal plane divacancies for neighboring $Si_{IIa}$ and $Si_{IIb}$ nuclei sites as shown in Fig. 3A formed two subgroups, with distinct properties (see Fig. 3E and Table S2,4).

4. The $k_3h_2$-ssf, $h_2k_2$-ssf, and bulk like *kh*-4H and *hk*-4H configurations.
5. The $k_1h_1$-ssf and $k_2k_1$-ssf basal configurations

In sum, the $k_2k_2$-ssf, $k_1h_1$-ssf, and $k_2k_1$-ssf divacancy configurations diverge from their defect counterparts within the quantum well's influence.

These results show that a single 1FSF(3,2) Frank-type stacking fault in 4H-SiC has a marked perturbative effect on the spin density and derived properties of a large subset of the divacancy configurations in its region. Indeed, from the nine near fault configurations considered, three (one axial, and two basal plane) divacancies diverge from their bulk configurations. Interestingly, while the $k_2k_2$-ssf, $k_1h_1$-ssf, and $k_2k_1$-ssf configurations show differing hyperfine splittings and ZFS constants (Fig. 3B-F and Table S1-4) as compared to their unperturbed bulk-like counterparts, they maintain their sought-after magneto-optical and photoluminescence (PL) characteristics (see Table S1-5). Critically, the stacking fault also lowers the local conduction band minimum by ~0.24 eV, while the local charge defect states do not shift towards the valence band maximum.

Coincidentally, this slight change to the local band-gap could result in a dramatic improvement in the charge dynamics. Divacancy defects in SiC have a zero-phonon line (ZPL) emission around 1.1 eV and generally require ~1.3 eV to repump[16]. Lowering the local band-gap by 0.2 eV means that even at its lowest energy, the off-resonant pump laser can reset the charge state, resulting in significant improvements to charge stability. Therefore, the stacking fault opens pathways for the divacancy dark state (VV$^-$) to more readily repopulate the bright state (VV$^0$) [16] via excitation laser and single-photon ionization as depicted in Fig. 1D and Fig. S2, stabilizing the qubit.

These results provide a means by which to interpret recent optically detected magnetic resonance (ODMR) [25] and PL [18] studies in 4H-SiC that reported unexpected divacancy-like color centers, labeled as PL5-PL7 [7,25,26] and PL5'-PL6' [18], that cannot be simply assigned to the known structural bulk 4H-SiC configurations (*hh, kk, hk, kh*) [38]. In many respects, the optical properties of these anomalous color centers observed in commercial 4H-SiC display properties of bulk divacancy configurations. However, unlike the known bulk divacancy centers[38], PL5-7 show several outstanding characteristics, including stability during both room-temperature ODMR measurements[7,25] and photo-ionization studies[16]. Additionally, the PL intensity of these color centers is unaffected by the introduction of an appropriate re-pump laser[16]. This is in stark contrast to known bulk divacancy configurations that typically exhibit a many-orders-of-magnitude change in their photoluminescence signal[16].

This charge stability points to the fact that the PL5-7 defect spin is preferentially in its bright charge state under continuous off-resonant excitation, much like the predicted behavior of near-stacking fault divacancy configurations under excitation laser re-pump conditions. However, point defect-stacking fault complexes are rarely reported in experimental work on high quality crystalline SiC samples. For the assignment of PL5-7 to the distinct $k_2k_2$-ssf, $k_1h_1$-ssf, and $k_2k_1$-ssf configurations to be plausible, the commercially available 4H-SiC wafers which have been found to commonly feature PL5-PL7 [25] and PL5'-PL6' [18] signatures should harbor stacking faults as modeled above.

In order to validate this, we investigated multiple commercial sources of 0001-oriented (in Hexagonal Miller-Bravais indices notation) 4H SiC wafers by means of X-ray diffraction at Sector 12 ID-D of the Advanced Photon Source (see Methods for technical details). Crystallographically, 1FSF(3,2) Frank-type stacking faults are synonymous with the presence of 6H polytypic inclusions in a bulk 4H crystal. We thus performed two types of X-ray measurements to isolate such 6H inclusions within several commercially available 4H crystals: 1) symmetric diffraction measurements were made of the crystal truncation rod in the vicinity of the 4H-(0004) and 4H-(0008) Bragg peaks to look for evidence of scattering contributions from a minority phase of 6H; and 2) non-symmetric diffraction measurements were performed at a scattering condition where 4H diffraction is forbidden, and 6H diffraction is allowed. Both measurements were designed to identify the presence of 6H inclusions that share lattice orientation with the 4H host, which would likely be the case for any 6H inclusion described by the 1FSF(3,2) stacking fault.

The results of the symmetric diffraction measurements are shown in Figures 4A-B for one of the substrates tested. They show that the very strongly diffracting 4H-(0004) and 4H-(0008) Bragg peaks both feature weakly-diffracting shoulders indicative of the presence of a second phase in the crystal with a slightly larger $c$-axis interplanar spacing. Fitting of these peaks indicate that this minority phase has a $c$-axis interplanar spacing that is 0.05 ±0.01 % larger than that of bulk 4H-SiC. Further evidence of the presence of 6H inclusions is shown in the results of the non-symmetric x-ray diffraction measurement (Figure 4C-D). The observation of 6H-(10$\bar{1}$4) and 6H-(10$\bar{1}$8) Bragg peaks where 4H diffraction is forbidden unambiguously demonstrate the presence of inclusions of the 6H polytype in the 4H host crystal. These results confirm that 1FSF(3,2) Frank-type stacking faults that result in 6H polytype inclusions in 4H-SiC are readily observed in typical commercial wafers. Specifically, from the data shown in Figure 4, we estimate a 0.128

±0.012 volume percent of 6H inclusions in that particular 4H wafer. Similarly, the synchrotron diffraction measurements we performed revealed the same fingerprints of 6H inclusions in all of the 4H materials investigated, with varying absolute quantities of volumetric fraction.

Though their origins and locations cannot be deduced from our x-ray measurements, these stacking faults have been reported to form preferentially near the surface in as-purchased SiC wafers, possibly induced by face polishing[40]. In this case, point defect-stacking fault complexes would be expected in higher concentration in the near surface region of any commercially processes SiC sample. This prediction was verified via ODMR measurements in an inhomogeneous magnetic field to locate different axial divacancy color centers in the sample (see Fig. S1A-B). Indeed, it can be confirmed that the spatial distribution of the PL6 center is significantly different from that of the *hh*, *kk*, *hk*, and *kh* (PL1-4) divacancies as a function of substrate depth. The *hh*, *kk*, *hk*, and *kh* divacancies were found to be uniformly distributed through the sample, whereas the PL6 qubits were concentrated near the surfaces, matching the expected distribution of stacking faults.

Building on these results, we propose that the as-yet-unidentified divacancy-like color centers PL5-7 observed in commercial 4H-SiC are related to specific point defect structures positioned in the vicinity of a stacking fault that acts as a local quantum well within 4H-SiC. Specifically, based on our modeling of hyperfine splitting, our experimental measurement of 6H inclusions in 4H bulk, and our observation of surface dependence of PL6 signatures, we stipulate that the $k_1h_1$-ssf, $k_2k_2$-ssf, and $k_2k_1$-ssf configurations account for the previously unattributed PL5-7 signatures. The calculated hyperfine, zero-field-splitting, and ZPL magneto-optical parameters obtained for the $k_2k_2$-ssf configuration agree well with the experimental data reported for PL6 divacancy related qubit (see Fig. 3, and Supplementary Materials). Consequently, we assign the $k_2k_2$-ssf stacking fault-divacancy configuration to the PL6 signature and the experimentally observed room temperature stability of the PL6 center is attributed to the quantum-well stabilization mechanism discussed above and depicted in Fig. 1D.

Compared to the PL6 center, the basal-plane oriented PL5 and PL7 centers have more sparsely available experimental data. Nonetheless, they have several important experimentally observed similarities to the PL6 center, including similar optical- and spin-transition energies, a nearly identical ODMR stability with temperature, and a similar ODMR variation across wafers in multi-wafer studies[25]. These studies, in combination with our models, provide an understanding that the PL5 and PL7 centers are basal-plane-oriented homologues of the c-axis-oriented PL6 center. Although we cannot uniquely distinguish each center, we assign the PL5 and PL7 centers to the k1h1-ssf and k2k1-ssf subgroup configurations of 6H polytypic inclusions in 4H-SiC.

**Discussion**

We demonstrated that the quantum well of a stacking fault in 4H-SiC can give rise to a point defect qubit stabilization mechanism, without the application of a re-pumping laser[16]. Furthermore, we found an association between the Frank-type stacking faults in commercial 4H-SiC materials and the presence of the PL5-7 defect qubits. In this context, we were able to identify the PL5-7 ($k_1h_1$-ssf, $k_2k_2$-ssf, and $k_2k_1$-ssf) room-temperature qubits as divacancies in stacking-fault structures.

Our results demonstrate the prospect of utilizing proximity to stacking faults as a means to engineer and tailor the properties of divacancy complexes in SiC. Such an approach may provide a practical route to synthesize qubits with enhanced photoionization and room temperature stability, given that stacking faults are common low-energy defects in systems that display prolific polytypism. In SiC, stacking faults have been reported to form as a consequence of different fabrication treatments, such as doping[29], implantation[41], irradiation[42], annealing[29], plastic deformation[43], surface polishing[40], and even due to optical irradiation[44]. Any of these approaches can be harnessed to intentionally introduce stacking faults in specific locations of a 4H host crystal. Finally, recently developments in 3D engineering of defect creation[45] make it possible to synthesize near-stacking fault divacancies.

The study of point defects embedded in extended defects[46] presents unique challenges and complexities as compared to that of point defects in single crystal bulk materials[27]. However, these structures also provide opportunities to broaden the palette of technologically applicable point defect qubits with superior functionality. We explore this general approach through the specific example of a single stacking fault in 4H-SiC that stabilizes divacancy qubits in its vicinity. Stacking faults and other means of generating local quantum wells are readily available in many semiconductor systems that are suitable hosts for point defect qubits. For example, diamond, an important material for optically addressable spin qubits, also contains stacking faults[47]. Moreover, our underlying theory of defect qubits in quantum wells could be generalized to many semiconductor systems[48].

Thus, tuning the optoelectronic properties of point defects via local quantum wells could be an important strategy for discovering a large and robust class of new spin qubits.

**Methods**

In our first principles density functional theory (DFT) calculations, we use Vienna Ab initio Simulation Package (VASP)[49], a plane wave basis set of 420 eV, and the projector augmented wave method (PAW)[50] to describe electronic states of different defective 4H-SiC supercells. Perdew-Burke-Ernzerhof (PBE)[51] and Heyd-Scuseria-Ernzerhof (HSE06)[52] exchange correlation functionals are employed to include exchange-correlation effects of the many electron system at different levels of approximation. To preserve periodicity, two single stacking faults are included and placed 27.8 Å away from each other in our supercell models, which were fixed to have a total size of 55.7 Å in the $c$ direction. The large axial size of the supercell allows us to calculate and compare near-stacking fault as well as bulk-like divacancies using the same model by tuning the proximity of the defect to the stacking fault plane. Following the guidelines of Ref. [38,53], the basal planar size as well as the k-point grid density are optimized for all the magneto-optical parameters calculated in this study. To obtain the most accurate ground state hyperfine tensors of first neighbor $^{13}$C and second neighbor $^{29}$Si nuclei, the HSE06 functional was used on a PBE relaxed supercell of 704 atoms with 3×3×1 k-point sampling. To obtain the ground state ZFS, we used a 1584 atom supercell with 2×2×1 k-point set, PBE Kohn-Sham wavefunctions, and our in-house implementation for the ZFS tensor calculation[54]. In our computational study we concentrated on the most reliable ground state hyperfine and ZFS data, and we also calculated the ZPL energies (see Supplementary Materials).

For the band structure calculations, we used the HSE06[52] hybrid functional, a 420 eV plane wave cut-off energy, and 12×12×6 and 12×12×1 k-point sets for 4H and the stacking fault model, respectively.

The XRD experiments were carried out at beamline 12 ID-D of the Advanced Photon Source (APS), using a 6-axis goniometer specifically developed for high-resolution crystal truncation rod x-ray studies. The sample that was studied was a (0001)-surface-oriented 4H-SiC high purity semi-insulating wafer purchased from Cree Inc. An x-ray beam with a photon energy of 20 keV was used for all diffraction experiments. Two general diffraction geometries were used depending on the diffraction conditions targeted: symmetric (specular) and non-symmetric (off-specular). In specular x-ray diffraction, the incident and exit angle of the incoming and diffracted beams are equal, such that the photon momentum transfer (or q-vector) is oriented along the surface normal. Since the (0001) crystallographic planes were parallel to the wafer surface, this specular diffraction geometry gave access to information about the (0001) layer spacing in the crystal. The non-specular x-ray diffraction geometry was utilized to reach positions along an off-specular crystal truncation rod where 6H SiC diffraction was allowed, but 4H-SiC diffraction was forbidden. Both geometries involved measurements of a crystal truncation rod along the out-of-plane c-axis crystallographic direction, known as $L$-scans. These were angular sweeps of appropriate goniometer axes that spanned equal spaced steps along the crystal truncation rod of interest, $(000L)$ for specular and $(10\bar{1}L)$ for off-specular. The units of L in the figure are given as reciprocal lattice units of the 4H-SiC lattice. The (0004)/ (0008) 4H and 6H diffraction doublets were fitted using Voigt doublets that accounted for consistent shape factors and intensity ratios within and throughout both diffraction windows, respectively. The volumetric fraction was approximated by using the ratio of areas for the diffraction doublets.

**References**

1. Wrachtrup, J. & Jelezko, F. Processing quantum information in diamond. *J. Phys.-Condens. Matter* **18**, S807–S824 (2006).

2. Weber, J. R. *et al.* Quantum computing with defects. *PNAS* **107**, 8513–8518 (2010).

3. Aharonovich, I., Englund, D. & Toth, M. Solid-state single-photon emitters. *Nat. Photonics* **10**, 631 (2016).

4. Maze, J. R. *et al.* Nanoscale magnetic sensing with an individual electronic spin in diamond. *Nature* **455**, 644–647 (2008).

5. Kucsko, G. *et al.* Nanometre-scale thermometry in a living cell. *Nature* **500**, 54 (2013).

6. Aslam, N. *et al.* Nanoscale nuclear magnetic resonance with chemical resolution. *Science* **357**, 67–71 (2017).

7. Koehl, W. F., Buckley, B. B., Heremans, F. J., Calusine, G. & Awschalom, D. D. Room temperature coherent control of defect spin qubits in silicon carbide. *Nature* **479**, 84 (2011).

8. Christle, D. J. *et al.* Isolated electron spins in silicon carbide with millisecond coherence times. *Nat Mater* **14**, 160–163 (2015).

9. Widmann, M. *et al.* Coherent control of single spins in silicon carbide at room temperature. *Nat Mater* **14**, 164–168 (2015).

10. Kraus, H. *et al.* Room-temperature quantum microwave emitters based on spin defects in silicon carbide. *Nat Phys* **10**, 157–162 (2014).

11. Whiteley, S. J. *et al.* Spin–phonon interactions in silicon carbide addressed by Gaussian acoustics. *Nat. Phys.* **15**, 490–495 (2019).

12. Jelezko, F. & Wrachtrup, J. Single defect centres in diamond: A review. *Phys. Status Solidi A* **203**, 3207–3225 (2006).

13. Awschalom, D. D., Bassett, L. C., Dzurak, A. S., Hu, E. L. & Petta, J. R. Quantum Spintronics: Engineering and Manipulating Atom-Like Spins in Semiconductors. *Science* **339**, 1174–1179 (2013).

14. Fu, K.-M. C., Santori, C., Barclay, P. E. & Beausoleil, R. G. Conversion of neutral nitrogen-vacancy centers to negatively charged nitrogen-vacancy centers through selective oxidation. *Appl. Phys. Lett.* **96**, 121907 (2010).

15. Aslam, N., Waldherr, G., Neumann, P., Jelezko, F. & Wrachtrup, J. Photo-induced ionization dynamics of the nitrogen vacancy defect in diamond investigated by single-shot charge state detection. *New J. Phys.* **15**, 013064 (2013).

16. Wolfowicz, G. *et al.* Optical charge state control of spin defects in 4H-SiC. *Nat. Commun.* **8**, 1876 (2017).

17. Golter, D. A. & Lai, C. W. Optical switching of defect charge states in 4H-SiC. *Sci. Rep.* **7**, 13406 (2017).

18. Magnusson, B. *et al.* Excitation properties of the divacancy in 4H-SiC. *Phys Rev B* **98**, 195202 (2018).

19. Dhomkar, S., Zangara, P. R., Henshaw, J. & Meriles, C. A. On-Demand Generation of Neutral and Negatively Charged Silicon-Vacancy Centers in Diamond. *Phys Rev Lett* **120**, 117401 (2018).

20. Balasubramanian, G. *et al.* Ultralong spin coherence time in isotopically engineered diamond. *Nat. Mater* **8**, 383 (2009).

21. Buckley, B. B., Fuchs, G. D., Bassett, L. C. & Awschalom, D. D. Spin-Light Coherence for Single-Spin Measurement and Control in Diamond. *Science* **330**, 1212 (2010).

22. Anderson, C. P. *et al.* Electrical and optical control of single spins integrated in scalable semiconductor devices. *ArXiv190608328 Cond-Mat Physicsquant-Ph* (2019).

23. Riedel, D. *et al.* Resonant Addressing and Manipulation of Silicon Vacancy Qubits in Silicon Carbide. *Phys Rev Lett* **109**, 226402 (2012).

24. Tsuchida, H., Kamata, I. & Nagano, M. Formation of basal plane Frank-type faults in 4H-SiC epitaxial growth. *J. Cryst. Growth* **310**, 757–765 (2008).

25. Falk, A. L. *et al.* Polytype control of spin qubits in silicon carbide. *Nat. Commun* **4**, 1819 (2013).

26. Klimov, P. V., Falk, A. L., Christle, D. J., Dobrovitski, V. V. & Awschalom, D. D. Quantum entanglement at ambient conditions in a macroscopic solid-state spin ensemble. *Sci. Adv.* **1**, (2015).

27. Lohrmann, A. *et al.* Single-photon emitting diode in silicon carbide. *Nat Commun* **6**, 7783 (2015).

28. Sridhara, S. G., Carlsson, F. H. C., Bergman, J. P. & Janzén, E. Luminescence from stacking faults in 4H SiC. *Appl. Phys. Lett.* **79**, 3944–3946 (2001).

29. Kuhr, T. A., Liu, J., Chung, H. J., Skowronski, M. & Szmulowicz, F. Spontaneous formation of stacking faults in highly doped 4H–SiC during annealing. *J. Appl. Phys.* **92**, 5863–5871 (2002).

30. Izumi, S., Tsuchida, H., Kamata, I. & Tawara, T. Structural analysis and reduction of in-grown stacking faults in 4H–SiC epilayers. *Appl. Phys. Lett.* **86**, 202108 (2005).

31. Si, W., Dudley, M., Shuang Kong, H., Sumakeris, J. & Carter, C. Investigations of 3C-SiC inclusions in 4H-SiC epilayers on 4H-SiC single crystal substrates. *J. Electron. Mater.* **26**, 151–159 (1997).

32. Castelletto, S. *et al.* A Silicon Carbide Room-temperature Single-photon Source. *Nat Mater* **13**, 151–156 (2014).

33. Feng, G., Suda, J. & Kimoto, T. Characterization of stacking faults in 4H-SiC epilayers by room-temperature microphotoluminescence mapping. *Appl. Phys. Lett.* **92**, 221906 (2008).

34. Kamata, I., Zhang, X. & Tsuchida, H. Photoluminescence of Frank-type defects on the basal plane in 4H–SiC epilayers. *Appl. Phys. Lett.* **97**, 172107 (2010).

35. Chen, B. *et al.* Tuning minority-carrier lifetime through stacking fault defects: The case of polytypic SiC. *Appl. Phys. Lett.* **100**, 132108 (2012).

36. Benamara, M. *et al.* Structure of the carrot defect in 4H-SiC epitaxial layers. *Appl. Phys. Lett.* **86**, 021905 (2005).

37. Miao, K. C. *et al.* Electrically driven optical interferometry with spins in silicon carbide. arXiv: 1905.12780 (2019).

38. Davidsson, J. *et al.* First principles predictions of magneto-optical data for semiconductor point defect identification: the case of divacancy defects in 4H-SiC. *New J. Phys.* **20**, 023035 (2018).

39. Falk, A. L. *et al.* Optical Polarization of Nuclear Spins in Silicon Carbide. *Phys Rev Lett* **114**, 247603 (2015).

40. Ushio, S. *et al.* Formation of Double Stacking Faults from Polishing Scratches on 4H-SiC (0001) Substrate. in *Silicon Carbide and Related Materials 2013* **778**, 390–393 (Trans Tech Publications, 2014).

41. Chen, B. *et al.* Pinning of recombination-enhanced dislocation motion in 4H–SiC: Role of Cu and EH1 complex. *Appl. Phys. Lett.* **96**, 212110 (2010).

42. Regula, G. & Yakimov, E. B. Effect of low energy electron beam irradiation on Shockley partial dislocations bounding stacking faults introduced by plastic deformation in 4H-SiC in its brittle temperature range. *Superlattices Microstruct.* **99**, 226–230 (2016).

43. Pichaud, B., Regula, G. & Yakimov, E. B. Electrical and optical properties of stacking faults introduced by plastic deformation in 4H-SiC. *AIP Conf. Proc.* **1583**, 161–164 (2014).

44. Galeckas, A., Linnros, J. & Pirouz, P. Recombination-Induced Stacking Faults: Evidence for a General Mechanism in Hexagonal SiC. *Phys Rev Lett* **96**, 025502 (2006).

45. Ohshima, T. *et al.* Creation of silicon vacancy in silicon carbide by proton beam writing toward quantum sensing applications. *J. Phys. Appl. Phys.* **51**, 333002 (2018).

46. Xi, J., Liu, B., Zhang, Y. & Weber, W. J. Ab initio study of point defects near stacking faults in 3C-SiC. *Comput. Mater. Sci.* **123**, 131–138 (2016).

47. Zhu, W. Defects in Diamond. in *Diamond: Electronic Properties and Applications* (eds. Pan, L. S. & Kania, D. R.) 175–239 (Springer US, 1995). doi:10.1007/978-1-4615-2257-7_5

48. Zhou, Y. *et al.* Room temperature solid-state quantum emitters in the telecom range. *Sci. Adv.* **4**, eaar3580 (2018).

49. Kresse, G. & Hafner, J. *Ab initio* molecular-dynamics simulation of the liquid-metal–amorphous-semiconductor transition in germanium. *Phys Rev B* **49**, 14251–14269 (1994).

50. Blöchl, P. E. Projector augmented-wave method. *Phys Rev B* **50**, 17953–17979 (1994).

51. Perdew, J. P., Burke, K. & Ernzerhof, M. Generalized Gradient Approximation Made Simple. *Phys Rev Lett* **77**, 3865–3868 (1996).

52. Heyd, J., Scuseria, G. E. & Ernzerhof, M. Hybrid functionals based on a screened Coulomb potential. *J Chem Phys* **118**, 8207 (2003).

53. Ivády, V., Abrikosov, I. A. & Gali, A. First principles calculation of spin-related quantities for point defect qubit research. *Npj Comput. Mater.* **4**, 76 (2018).

54. Ivády, V., Simon, T., Maze, J. R., Abrikosov, I. A. & Gali, A. Pressure and temperature dependence of the zero-field splitting in the ground state of NV centers in diamond: A first-principles study. *Phys Rev B* **90**, 235205 (2014).

**Additional references in Supplementary Materials**

55. Bockstedte, M., Schütz, F., Garratt, T., Ivády, V., & Gali, A. Ab initio description of highly correlated states in defects for realizing quantum bits. npj Quantum Materials **3**, 31 (2018).

**Acknowledgments**

**General:** We thank Christopher P. Anderson, Gary Wolfowicz, and Sean Sullivan for careful reading of the manuscript. We also thank Angel Yanguas-Gil and Hua Zhou for experimental assistance.

**Funding:** VI acknowledges the support from the MTA Premium Postdoctoral Research Program. This work was financially supported by the Knut and Alice Wallenberg Foundation through WBSQD2 project (Grant No. 2018.0071). Support from the Swedish Government Strategic Research Areas in Materials Science on Functional Materials at Linköping University (Faculty Grant SFO-Mat-LiU No. 2009-00971) and the Swedish e-Science Centre (SeRC), the Swedish Research Council (VR 2016-04068), and the Carl-Trygger Stiftelse för Vetenskaplig Forskning (CTS 15:339) is gratefully acknowledged. Analysis of the model Hamiltonians parameters was supported by the Ministry of Education and Science of the Russian Federation in the framework of Increase Competitiveness Program of NUST MISIS (No No K2-2019-001) implemented by a governmental decree dated 16 March 2013, No 211. AG acknowledges the Hungarian NKFIH grants No. KKP129866 of the National Excellence Program of Quantum-coherent materials project, No. 127902 of the EU QuantERA Nanospin project, EU H2020 Quantum Technology Flagship project ASTERIQS (Grant No. 820394), the NVKP project (Grant No. NVKP_16-1-2016-0043) as well as the National Quantum Technology Program (Grant No. 2017-1.2.1-NKP-2017-00001). The calculations were performed on resources provided by the Swedish National Infrastructure for Computing (SNIC 2016/1-528, SNIC 2017/11-8, SNIC 2018/3-625) at the National Supercomputer Centre (NSC) and by Linköping University (LiU-2015-00017-60). X-ray measurements were performed at the Advanced Photon Source, a U.S. Department of Energy (DOE) Office of Science User Facility operated for the DOE Office of Science by Argonne National Laboratory under Contract No. DE-AC02-06CH11357. N.D., S.O.H., M.V.H., F.J.H., and D.D.A. were supported by the DOE, Office of Basic Energy Sciences.

**Author contributions:** V.I. and N.D. wrote the manuscript with input from all authors. Calculations were performed by V.I. and J.D. with input from N.T.S., I.A.A., and A.G. Optical experiments were performed by A.L.F., P.V.K., and D.D.A. X-ray measurements were performed by N.D., S.J.W., S.O.H., M.V.H., and F.J.H. The computational and experimental results were analyzed with contributions from all authors.

**Competing interests:** The authors declare no competing interests.

**Figures**

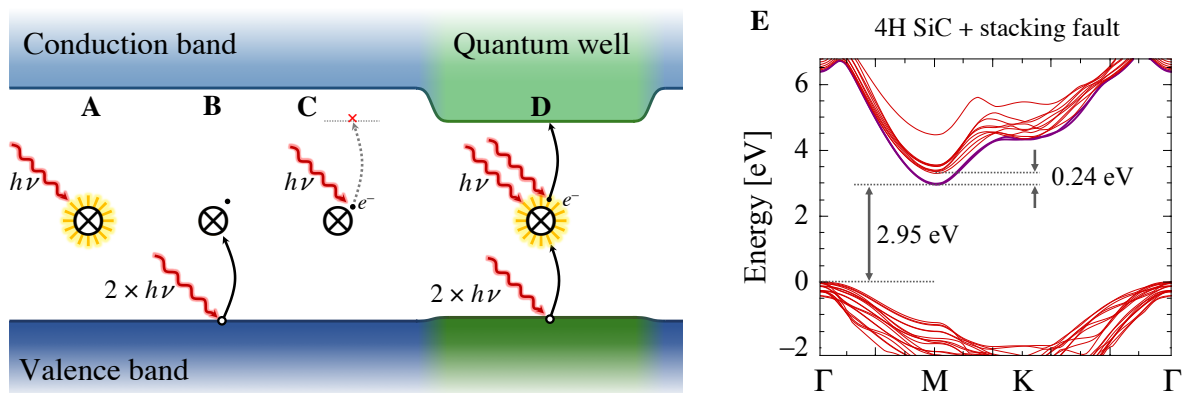

**Fig. 1. A color center with a quantum well stabilized bright state. A** A bright state of a color center under optical excitation. **B** These incident photons may ionize the defect and turn it into a dark state while also, **C**, not have sufficient energy to repopulate the bright state. **D** In a quantum well, however, the excitation laser can successfully re-pump the bright state. **E** Band structure of a defective 4H-SiC including a stacking fault. Red curves depict bulk-like conduction and valence band states in the basal plane of the hexagonal Brillouin zone. Purple curves show the stacking fault states that localized in $c$ direction and dispersive in the basal plane.

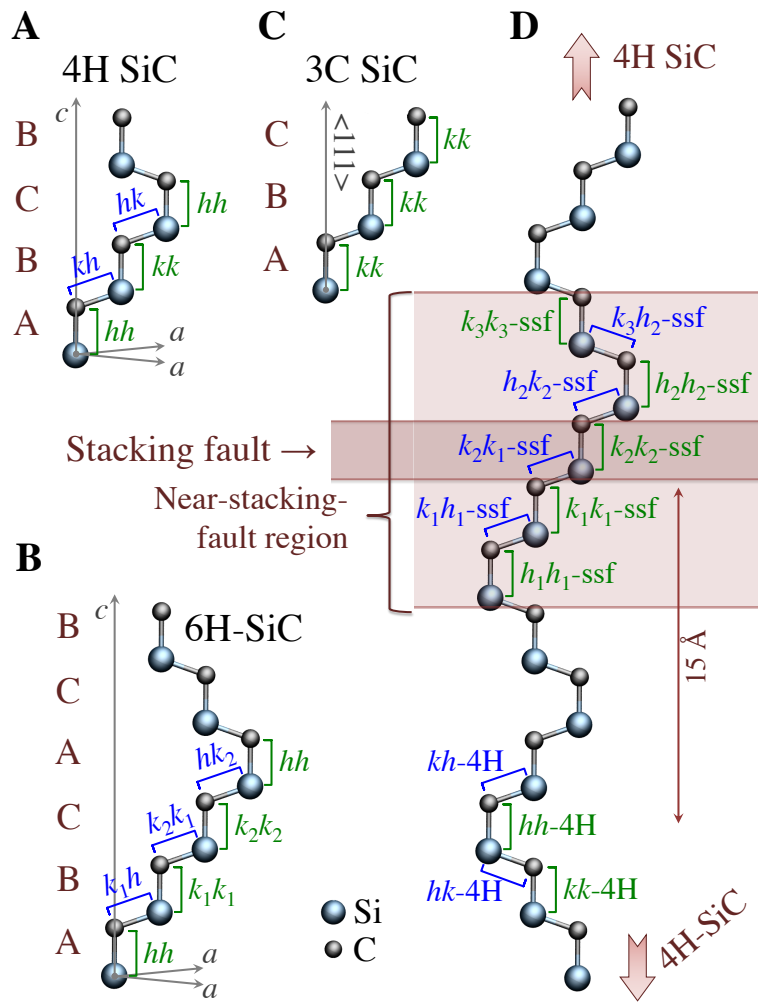

**Fig. 2. Primitive cells of common SiC polytypes and the structure of a single Frank-type stacking fault in 4H-SiC. A**, **B**, and **C** show the primitive cells, the stacking sequences, and the possible divacancy nonequivalent configurations in 4H, 6H, and 3C-SiC, respectively. Here, *h* and *k* stand for hexagonal-like and cubic-like environments of Si or C sites, respectively, and the double letters for the vacancy sites of the $V_{Si}$-$V_C$ divacancy pair defect configurations. **D** A single stacking fault in a cubic-like stacking order in 4H-SiC. The close vicinity of the stacking fault resembles the 6H stacking and thus it gives rise to 6H-like additional divacancy configurations in 4H-SiC. The $k_2k_2$-ssf configuration is assigned to PL6 room-temperature qubits.

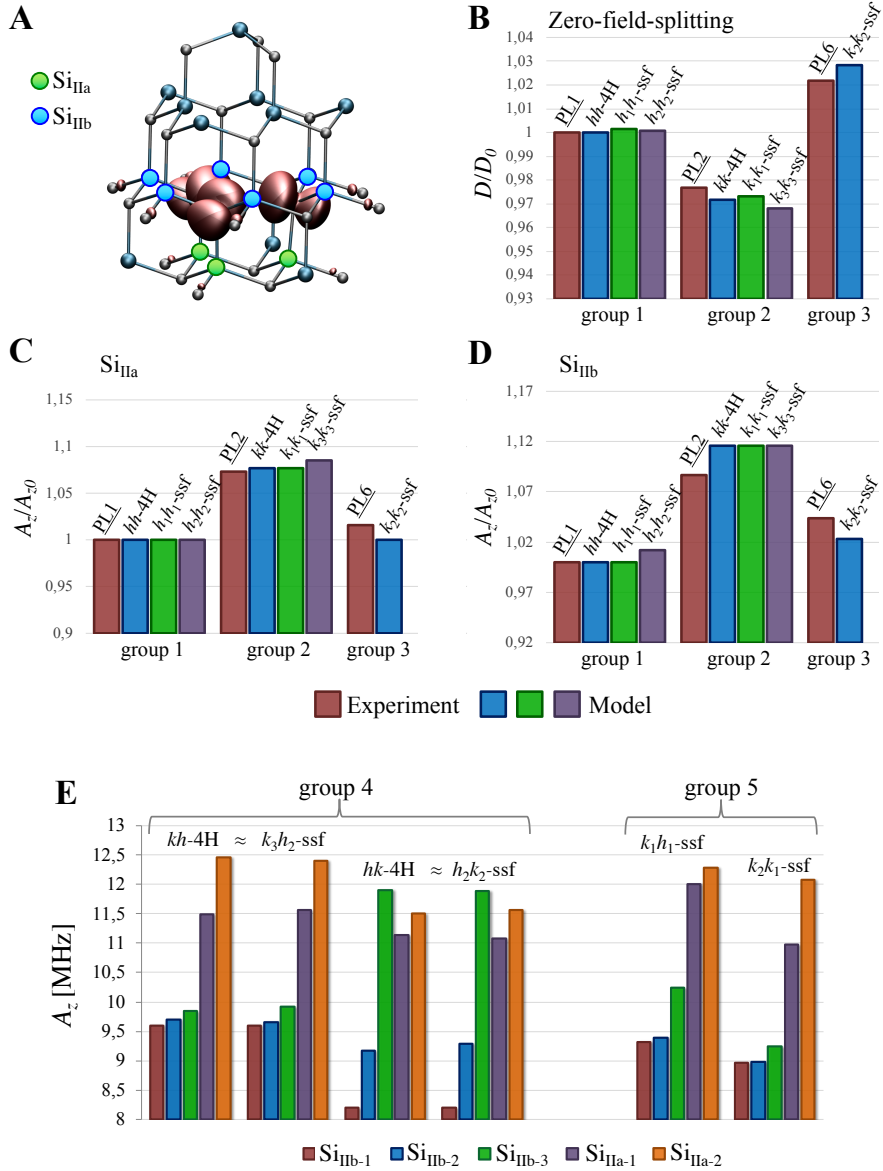

**Fig. 3 Hyperfine splitting $A_z$ and zero field splitting $D$ of divacancy configurations a defective 4H-SiC including a stacking fault.** **A** Spin density of an axial divacancy. Blue and green filled circles indicate the second neighbor silicon sites for which the hyperfine tensors were calculated. **B** The calculated and experimentally measured relative ground-state zero-field splitting parameter for each axial symmetric defect. $D_0$ is equal to the ZFS of *hh*-4H configuration and PL1 center in the calculation of the theoretical and experimental parameters, respectively. The experimental results were reported in Ref. [39]. **C**, **D** Relative

hyperfine parameters of $Si_{IIa}$ and $Si_{IIb}$ neighboring nuclei sites, which are depicted in **A**. $A_{z0}$ is equal to the hyperfine splitting of *hh*-4H configuration and PL1 center in the calculation of the theoretical and experimental parameters, respectively. **E** Hyperfine splitting of basal plane divacancy configurations. Due to the low symmetry of basal plane configurations, we distinguish symmetrically non-equivalent sites, $Si_{IIa-1}$ and $Si_{IIa-2}$, and $Si_{IIb-1}$, $Si_{IIb-2}$, and $Si_{IIb-3}$, in the neighbor shells of $Si_{IIa}$ and $Si_{IIb}$ shown in **A**.

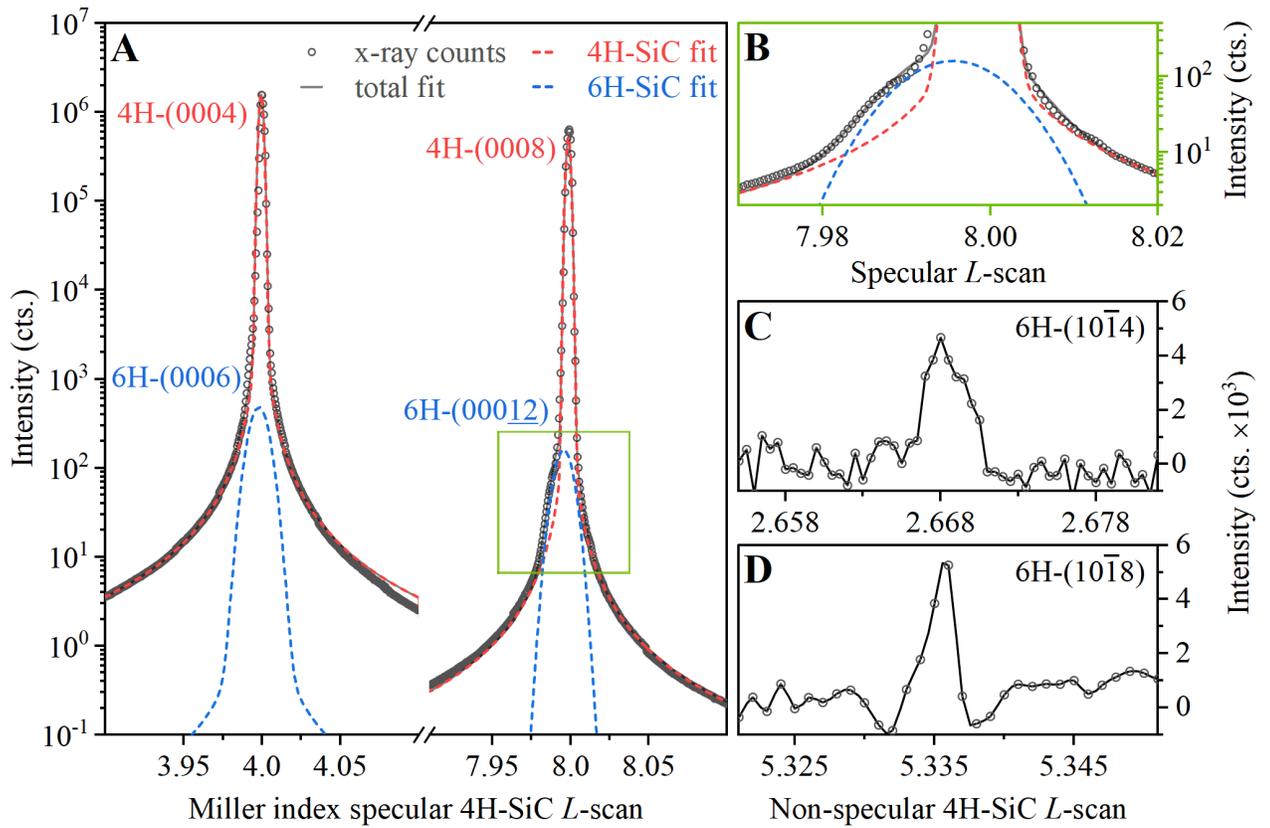

**Fig. 4. Synchrotron X-ray diffraction studies of polytypic 6H inclusions in commercially available 4H-SiC.** (**A**) Shows the (0004)/(0008) L-scan of 4H-SiC and contributions of the detected (0006)/(00012) contributions from 6H-SiC. We note the 4H and 6H components fitted simultaneously as fully inter-related doublets. (**B**) Zoom-in of the 4H-(0008) diffraction peak skew due to the presence of 6H-SiC polytypic inclusions. (**C** & **D**) The non-specular (10-14) & (10-18) peaks of 6H, respectively.